\documentclass[journal,twoside]{IEEEtran}

\usepackage{graphicx}
\usepackage[font=footnotesize, caption=false]{subfig}

\usepackage{flushend}

\makeatletter
\long\def\@makecaption#1#2{\ifx\@captype\@IEEEtablestring%
\footnotesize\begin{center}{\normalfont\footnotesize #1}\\
{\normalfont\footnotesize\scshape #2}\end{center}%
\@IEEEtablecaptionsepspace
\else
\@IEEEfigurecaptionsepspace
\setbox\@tempboxa\hbox{\normalfont\footnotesize {#1.}~~ #2}%
\ifdim \wd\@tempboxa >\hsize%
\setbox\@tempboxa\hbox{\normalfont\footnotesize {#1.}~~ }%
\parbox[t]{\hsize}{\normalfont\footnotesize \noindent\unhbox\@tempboxa#2}%
\else
\hbox to\hsize{\normalfont\footnotesize\hfil\box\@tempboxa\hfil}\fi\fi}
\makeatother

\makeatletter
\def\ps@headings{%
\def\@oddhead{\mbox{}\scriptsize\rightmark \hfil }%
\def\@evenhead{\scriptsize \hfil \leftmark\mbox{}}%
\def\@oddfoot{}%
\def\@evenfoot{}}
\makeatother

\makeatletter
\def\ps@IEEEtitlepagestyle{%
\def\@oddhead{\mbox{}\scriptsize\rightmark \hfil }%
\def\@evenhead{\scriptsize \hfil \leftmark\mbox{}}%
\def\@oddfoot{}%
\def\@evenfoot{}}
\makeatother

\begin{document}

\title{Statistical Topics Concerning Radiometer Theory}

%
\author{\IEEEauthorblockN{Todd R. Hunter\IEEEauthorrefmark{1}\IEEEauthorrefmark{2} and Robert Kimberk\IEEEauthorrefmark{2}}\\
\IEEEauthorblockA{\IEEEauthorrefmark{1}{\em NRAO, 520 Edgemont Rd, Charlottesville, VA, USA}}\\
\IEEEauthorblockA{\IEEEauthorrefmark{2}{\em Harvard-Smithsonian Center for Astrophysics, 60 Garden St., Cambridge, MA, USA}}\\
\IEEEauthorblockA{\IEEEauthorrefmark{1}Contact: thunter@nrao.edu, phone +1 434 244 6836}
}

\markboth{26th International Symposium on Space Terahertz Technology, Cambridge, MA, 16-18 March, 2015.}%
{26th International Symposium on Space Terahertz Technology, Cambridge, MA, 16-18 March, 2015.}

\maketitle

\begin{abstract}
We present a derivation of the radiometer equation based on the original references and fundamental statistical concepts. We then perform numerical simulations of white noise to illustrate the radiometer equation in action. Finally, we generate $1/f$ and $1/f^2$ noise, demonstrate that it is non-stationary, and use it to simulate the effect of gain fluctuations on radiometer performance.
\end{abstract}


\section{Introduction}

While attempting to understand noise in tube amplifiers, Johnson \cite{Johnson1928} discovered that ``Statistical fluctuation of electric charge exists in all conductors, producing random variation of potential between the end of the conductor.''  The National Institute of Standards and Technology (NIST) defines a radiometer as “a very sensitive receiver, typically with an antenna input, that is used to measure radiated electromagnetic power” \cite{NIST2008}.  Radiometer theory connects Johnson's and NIST's concepts and leads to the radiometer equation, which is of universal importance to radio and terahertz astronomy as well as atmospheric studies because it describes the ultimate sensitivity of radiometers \cite{Racette2005}.  Unfortunately, most astronomy textbook and lecture notes on the radiometer equation do not present the details of this relationship, and often provide only a cursory overview, which can be misleading depending on the background of the reader. In this work, we present a derivation of the radiometer equation and explore the effects of non-stationary noise via numerical simulations. 

\section{Derivation of the Radiometer Equation}

\subsection{Variables and constants}
In the definitions below, the typical physical units are given in parentheses.
\begin{itemize}
\item $A^2$ = instantaneous power (watt) = the observed power at one statistically independent observation.
\item $A$ = amplitude ($\sqrt{\rm{watt}}$) = $V/\sqrt{R}$ where $V$ = Voltage (volt), and $R$ = resistance (ohm).
\item $\beta$ = equivalent noise bandwidth (Hz) of the radiometer. For Gaussian white noise, the equivalent noise bandwidth has a uniform power spectral density and represents a finite spectral interval of Johnson noise. Bandwidth 
in this description will refer to baseband bandwidth.
\item $\tau$ = time interval (sec) during which the radiometer measures mean power
\item $P$ = mean power (watt).
\item $N$ = number of statistically independent observations.
\item $k$ = Boltzmann's constant = 1.38*10$^{-23}$ (joule/kelvin)
\item $T$ = blackbody temperature of source (kelvin).
\item $\mu$ = mean of a distribution.
\item $\sigma$ = standard deviation of a distribution.
\item $\sigma^2$ = variance = $\Sigma(X-\mu)^2/N$.
\item $\sigma_{\rm SE}$ = standard error of the mean.
\item $\sigma_{\rm P}$ = standard error of mean power.
\item $\sigma_{\rm T}$ = standard error of blackbody temperature.
\item $z$ = standard normal random variable. A normal random variable with $\mu$=0, and $\sigma$=1.
\item $z^2$ = chi-square random variable, one degree of freedom which has $\sigma=\sqrt{2}$, and $\mu$=1.
\end{itemize}

\subsection{Useful Rule}

If $c$ is a constant, and $z^2$ a chi-square random variable with one degree of freedom, then $cz^2$ is a gamma distributed random variable and the standard deviation is $\sqrt{2}c$.  A normal random variable may be expressed as $\sigma z + \mu$.

\subsection{Definition}
A sample is a set of observations of a random variable. Most of the statistics involved in this derivation are sample statistics. If population statistics are known they may be substituted were permissible. 
The standard error is the standard deviation of a sample statistic. The standard error of the mean is: 
$\sigma_{\rm SE} = \sigma/\sqrt{N}$
where $\sigma$ is the standard deviation of the random variable being averaged, and $N$ is the number of statistically independent samples used to generate the mean. The standard error of the mean is the standard deviation of many sample means.

\subsection{Assumptions}

$A$ is a normal random variable with standard deviation $\sigma$ and mean $\mu$=0. The point of reference of this derivation is the output terminals of the radiometer antenna or output terminals of a resistor.  The amplitude $A$ is measured at the output terminals.
$T$, $A$, and $P$ are statistically stationary, meaning that the statistical moments and autocorrelation are not functions of time.

\subsection{Derivation}

The radiometer equation is a function that determines the standard error of the mean of instantaneous power or temperature.
\begin{itemize}
\item $A = \sigma z$,   a statistically independent observation of a normal random variable with zero mean. Note that $A^2$=$\sigma^2z^2$.
\item $P = \Sigma A^2/N = \sigma^2$ \cite{Nyquist1928}.
\item mean power = $\sigma^2 = \Sigma (A-\mu)^2/N$ with $\mu=0$.
\item But also: $P=\Sigma \sigma^2 z^2 / N$ given $A=\sigma z + \mu$ with $\mu=0$.
\item $\sqrt{2}\sigma^2$ is the standard deviation of $A^2$, given $z^2$ with $\mu=1$ and $\sigma=\sqrt{2}$.
\item Given that $\sqrt{2}\sigma^2$ is the standard deviation of $\sigma^2z^2$ the standard error of the mean of $P$ is $\sigma_{\rm P} = \sqrt{2}\sigma^2 / \sqrt{N}$.  This is the standard error of variance of a normal random variable \cite{Fisher1930}.
\item The number of statistically independent samples in a signal of bandwidth $\beta$ and in $\tau$ time is: $N = 2\beta\tau$ \cite{Oliver1965}.
\item The standard error of the mean power can then be written as:
\begin{itemize}
\item $\sigma_{\rm P} = \sqrt{2}\sigma^2/\sqrt{N}$
\item $\sigma_{\rm P} = \sqrt{2}\sigma^2/\sqrt{2\beta\tau}$
\item $\sigma_{\rm P} = \sigma^2/\sqrt{\beta\tau}$
\end{itemize}
\item Given Johnson's \cite{Johnson1928} formula $P = kT\beta$, the standard error of temperature is:
$\sigma_{\rm T} = T/\sqrt{\beta\tau}$. 
\end{itemize}

\section{Numerical simulation of the radiometer equation}
\label{sim}
\begin{figure}[!t]
\centering
\includegraphics[width=3.4in,trim=30 0 90 390,clip]{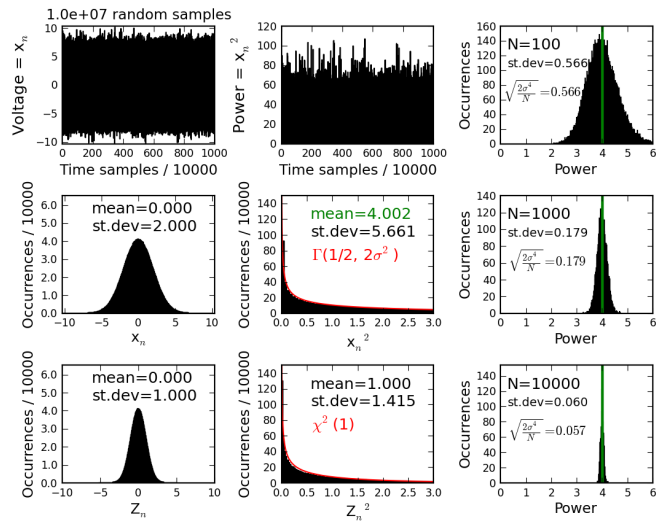}
\caption{Simulations of the radiometer equation with a stationary white noise signal, as described in section \S~\ref{sim}. The left column shows the amplitude time series along with its mean, standard deviation and PDFs.  The middle column shows the same things for the corresponding power signal. The right column shows how the PDF of the measured power becomes narrower around the true power (marked by the vertical green line) as the integration time increases, consistent with the prediction of the radiometer equation.}
\label{fig1}
\end{figure}

We have used the numerical python packages (numpy and scipy) to simulate and analyze  a radio frequency signal as a zero mean white noise datastream of 10 million points.  We set the standard deviation of $A=2$ in arbitrary units, resulting in a power of 4.   In Figure~\ref{fig1}, the left column shows the amplitude (voltage) signal vs. time, a histogram of its probability distribution function (PDF), and a histogram of its normalized PDF.   The middle column shows the same quantities for the amplitude squared (i.e. power) signal.  In this case, the power PDF is described by a Gamma function (shown by the red line), while the normalized power PDF is a chi-squared function with 1 degree of freedom.  In the right-most column, we show PDFs of the observed power which simulate a series of observations of increasing length. 
In the top right panel, we break the
datastream into $10^5$ samples each containing 100 observations. 
With only $N$=100 samples in each observation, the width of the PDF is large, showing that the signal to oise ratio (SNR) of one observation is only 7.07.  Increasing the number of samples to $N$=1000 and then 10000 shows the increasing precision on the measurement of the power level to a SNR of 22.3 then 70.2, consistent with the prediction of the radiometer equation.

\section{Numerical simulations with non-stationary noise}
\subsection{Importance of stationary vs. non-stationary noise}

The radiometer equation is based, in part, on the assumption of statistically stationary mean and variance of the Gaussian signal from the thermal source of radiation observed by the radiometer. This provides a good estimate of mean power and temperature. The estimate fails, at some level of precision, due to the time varying gain and noise temperature of the radiometer. Simple solutions to this issue can be found if the time variation is predicable, for instance a linear trend like one’s age, or a cyclostationary variation, like the average temperature in an hour of a day, which shows both diurnal and annual periodicity. Unfortunately the variation of radiometer gain and noise temperature are random. 

A common model for this behavior is a random variable whose power spectral density has a power law dependence. Johnson had to take a baseline measurement of the amplifier response with no signal at the input, and subtract this ``zero deflection'' from his measures of thermal noise. Dicke developed the Dicke switch method of radiometer gain calibration to overcome the gain variation \cite{Dicke}. Power law noise is typically found at the low frequency end of the power spectral density of an amplifier's output power \cite{Tsybulev2014}. The power is a function of frequency and may be represented by $P \propto f^{-\alpha}$ where $\alpha$ is slope of the power spectral density. 

\begin{figure}[!t]
\centering
\includegraphics[width=3.4in,trim=60 120 20 270,clip]{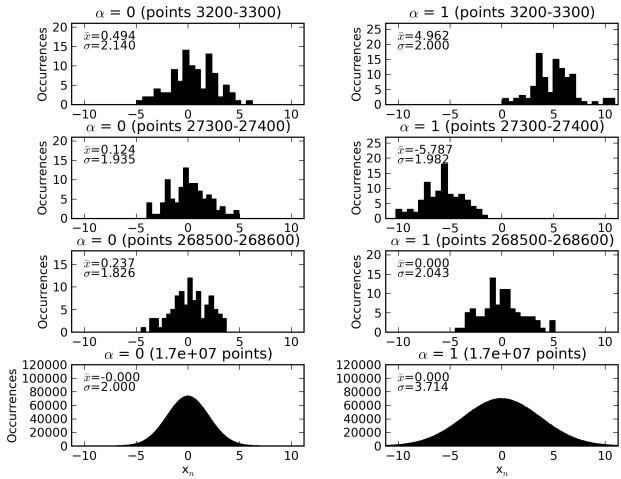}
\caption{The left column shows a stationary datastream – the mean and variance do not change with time.  The right column shows a non-stationary datastream in which 100 consecutive samples can sometimes be all positive (first row), or all negative (second row) or be centered at zero (third row).   Although both datastreams produce a Gaussian PDF when viewed in aggregate (bottom row), the variance diverges with time when $\alpha > 0$, as shown in Fig.~\ref{diverge}.}
\label{fig2}
\end{figure}

\subsection{Generating $1/f$ noise}
To simulate $1/f^{\alpha}$ noise with different values of $\alpha$, we use the C code of Paul Bourke \footnote{See {http://paulbourke.net/fractals/noise}}, which creates noise data by the ``fractional Brownian motion (fBm)'' method.  The fBm method involves creating frequency components which have a magnitude that is generated from a Gaussian white process and scaled by the appropriate power of $\alpha$ (the phase is uniformly distributed on $[0,2\pi]$) and then inverse Fourier transformed back into the time domain.  Figure~3 shows that the power spectrum for the resulting
$\alpha=1$ is indeed $1/f$.

\begin{figure}[!t]
\centering
\includegraphics[width=1.7in,trim=0 0 10 300,clip]{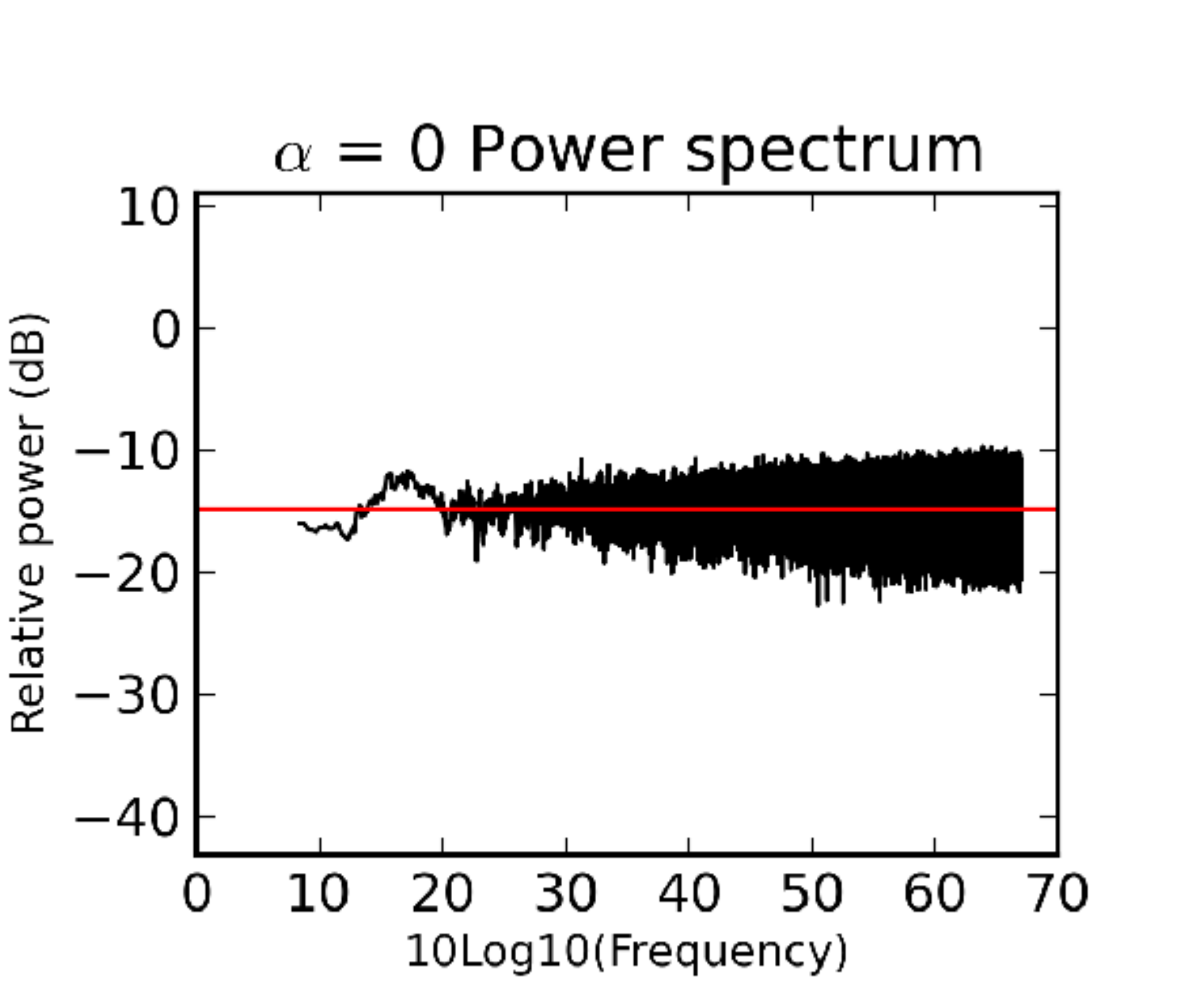}
\includegraphics[width=1.7in,trim=0 0 10 300,clip]{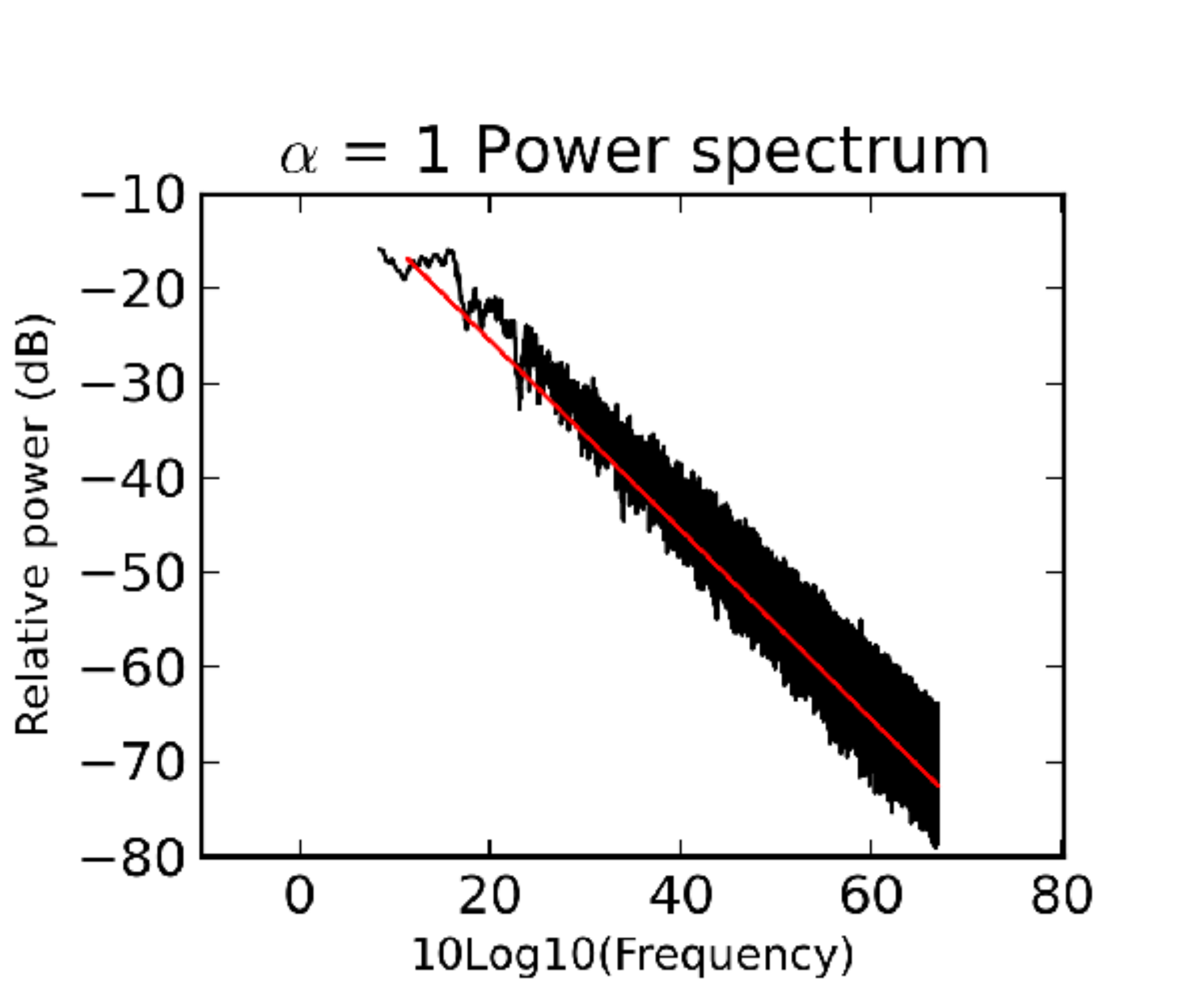}
\caption{Power spectrum of white noise ($\alpha=0$, left panel) compared to $\alpha=1$ noise (right panel)  The red line has a slope of zero and -1, respectively.}
\label{powerspectra}
\end{figure}

One feature of noise with $\alpha>0$ is that the variance diverges with time, in contrast to white noise whose variance approaches a constant value (Figure~4)

\begin{figure}[!t]
\centering
\includegraphics[width=2.8in,trim=60 160 55 200,clip]{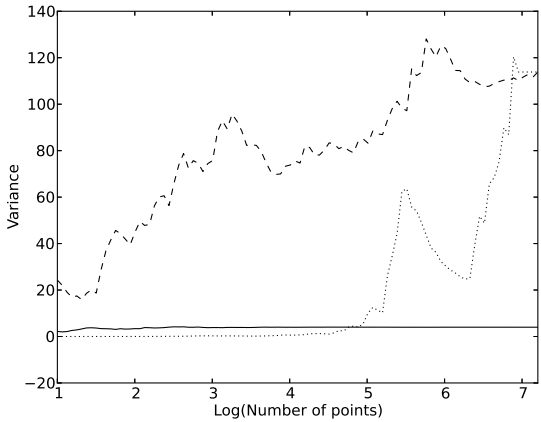}
\caption{Variance of a signal vs. the logarithm of the total time examined for three different random signals: $\alpha=0$ (solid line), $\alpha=1$ (dashed line), $\alpha=2$ (dotted line).  Signals with $\alpha > 0$ diverge with time.}
\label{diverge}
\end{figure}

In subsequent figures, we show how this noise differs from Gaussian white noise in a number of different representations.  In Figure~5, we show the structure of a portion the time-series, which exhibits a quasi-periodicity.

\begin{figure}[!t]
\centering
\includegraphics[width=1.7in,trim=0 0 150 360,clip]{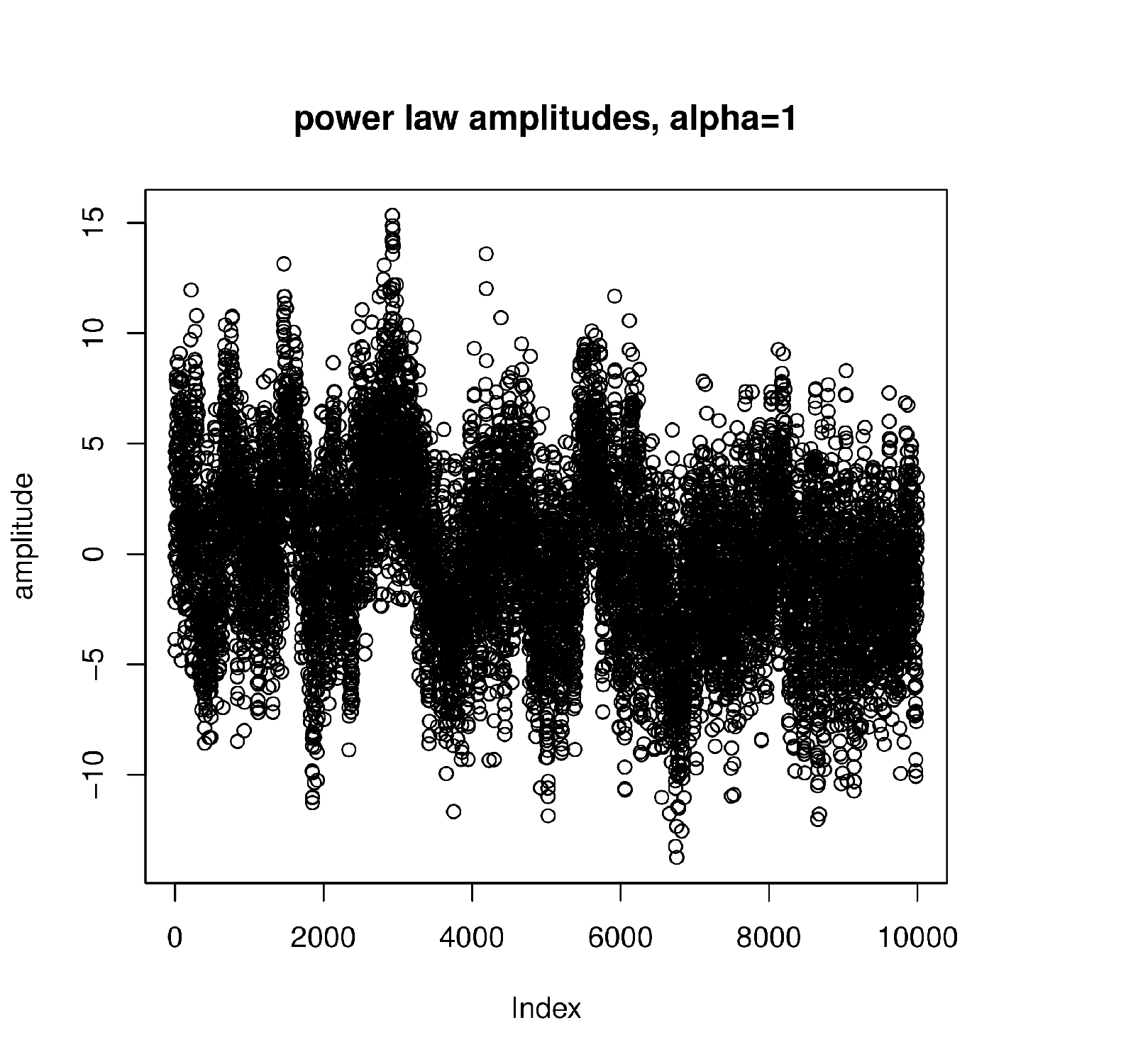}
\includegraphics[width=1.7in,trim=0 0 150 360,clip]{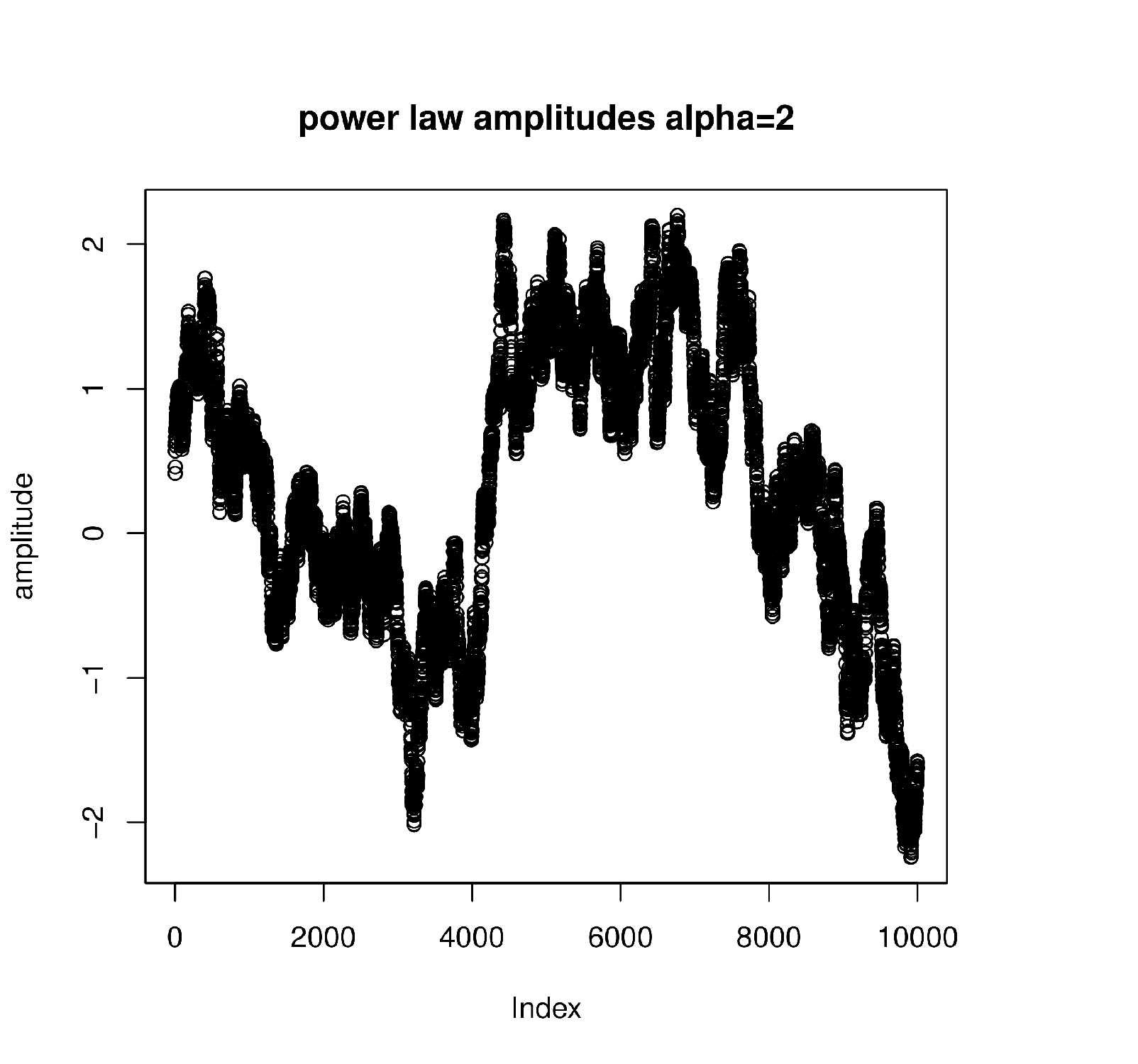}
\caption{Plots of ten thousand power law values vs. time for a signal 
with $\alpha=1$ (left panel) and $\alpha=2$ (right panel) show a 
periodicity which arises from a low frequency component of the noise. 
If we were to zoom into the plot, an image with similar features, 
though decreasing amplitudes would appear.}
\label{vstime}
\end{figure}

The auto-correlation functions (ACFs) are compared in Figure~6.  As expected, the ACF of white noise has signal only in the zero lag, and is zero at all other frequencies.  In costrast, a power law random variable displays a decreasing correlation with lag, and the details change with time.  In other words, the ACF changes when different independent samples are examined.

\begin{figure}[!t]
\centering
\includegraphics[width=3.4in,trim=70 150 10 250,clip]{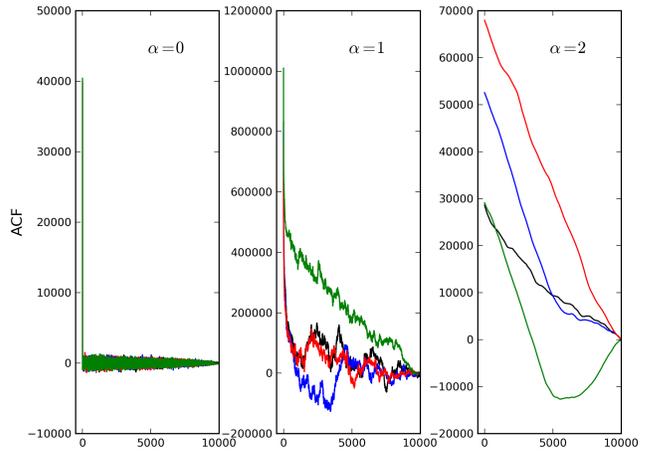}
\caption{Graphs of the autocorrelation function (ACF) of three random variables.  Left panel: white noise ($\alpha=0$); Center panel: power law noise with $\alpha=1$; right panel: power law noise with $\alpha=2$.  Different time samples are shown in different colors.  Note that the ACF changes with time for $\alpha>0$.}
\label{acfs}
\end{figure}

In Figure~7, we show the Allan variance of a signal which is primarily white noise but has a portion of $\alpha > 0$ noise added to it.  In Figure~8, we show how the Quantile-Quantile (QQ) relation for the $\alpha=1$ and $\alpha=2$ signals compares to that of a uniform distribution and a Cauchy distribution.

\begin{figure}[!t]
\centering
\includegraphics[width=2.3in,trim=5 16 200 420,clip]{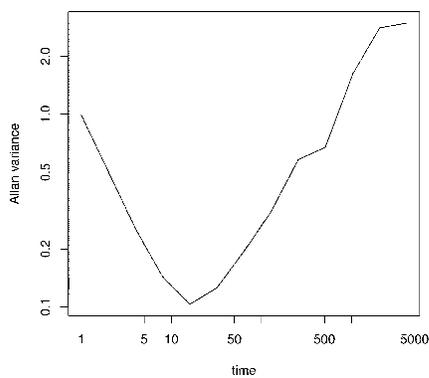}
\caption{Allan variance of a signal composed of the sum of white noise plus some $\alpha=2$ noise.}  
\label{allan}
\end{figure}

\begin{figure}[!t]
\centering
\includegraphics[width=1.44in,trim=5 18 200 420,clip]{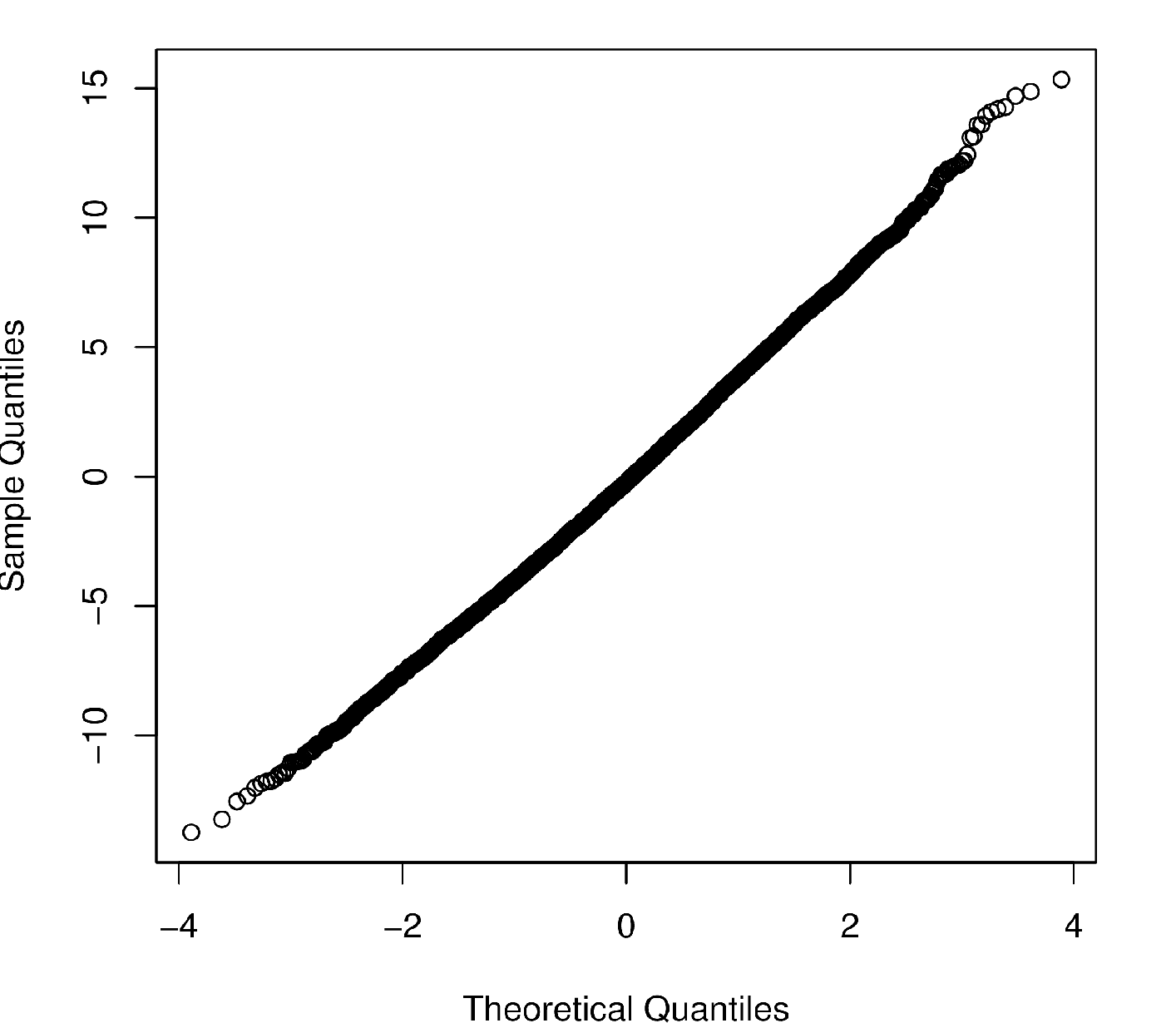}
\includegraphics[width=1.44in,trim=5 18 200 420,clip]{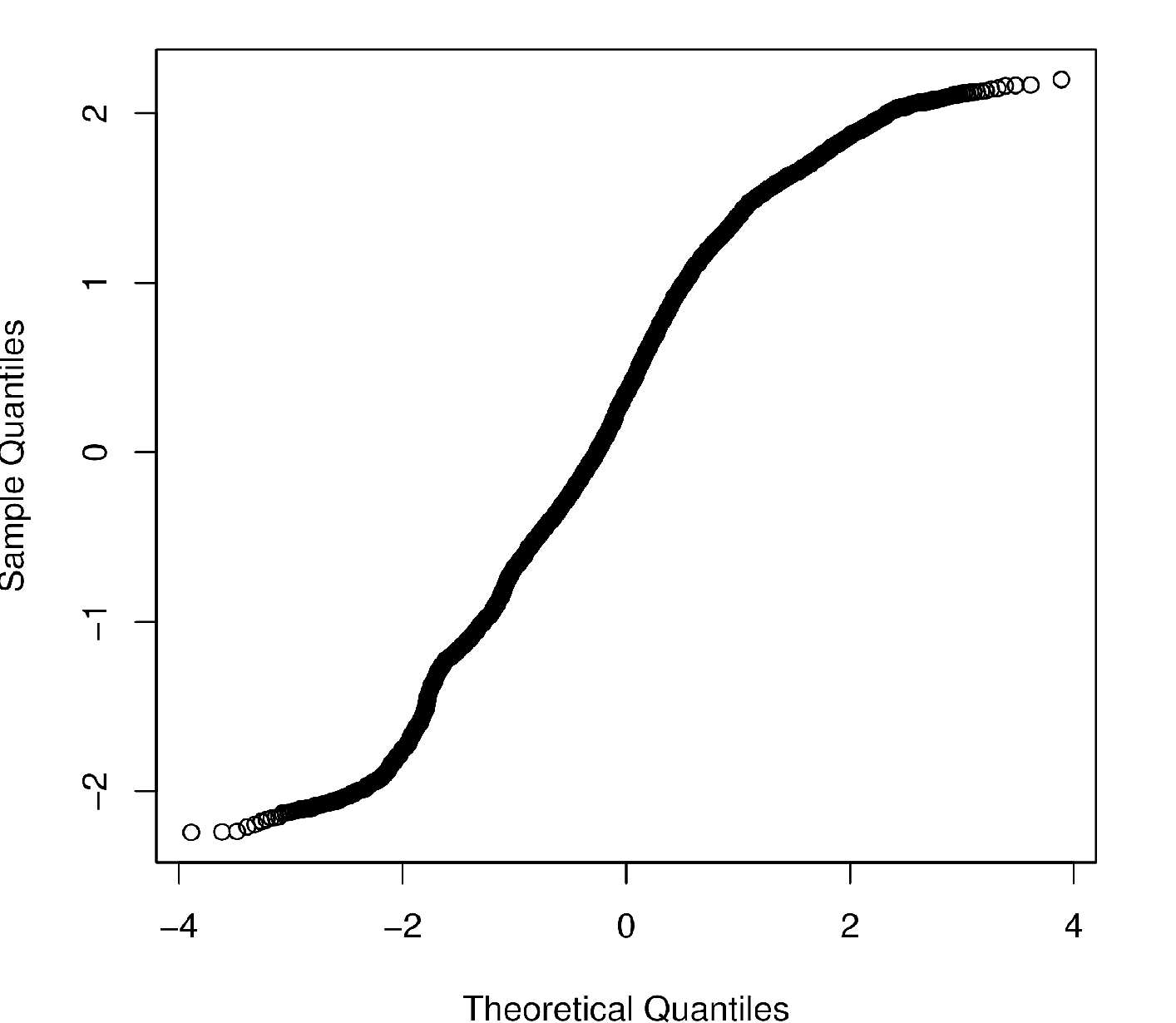}\\
\includegraphics[width=1.44in,trim=5 18 200 420,clip]{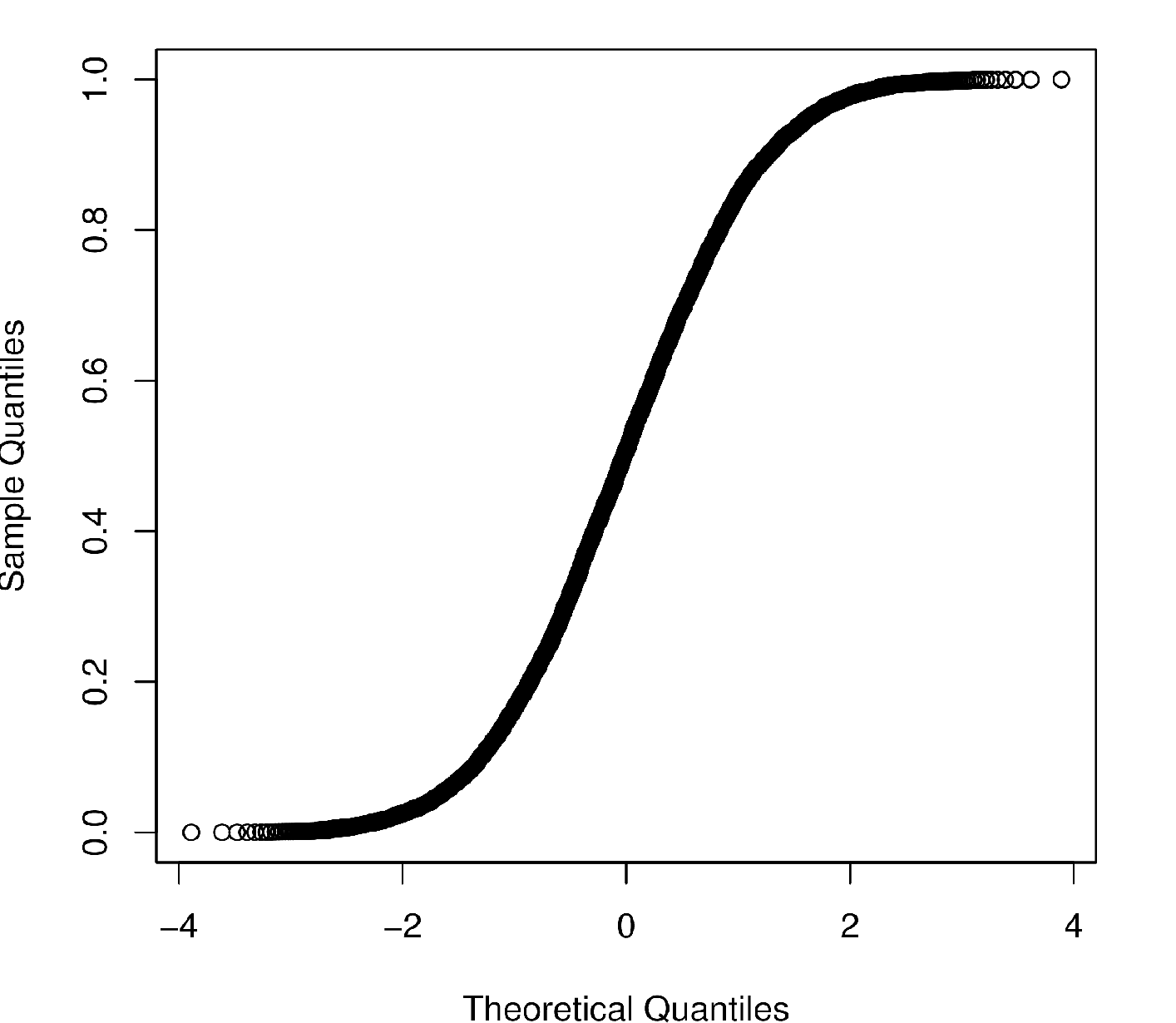}
\includegraphics[width=1.44in,trim=5 18 200 420,clip]{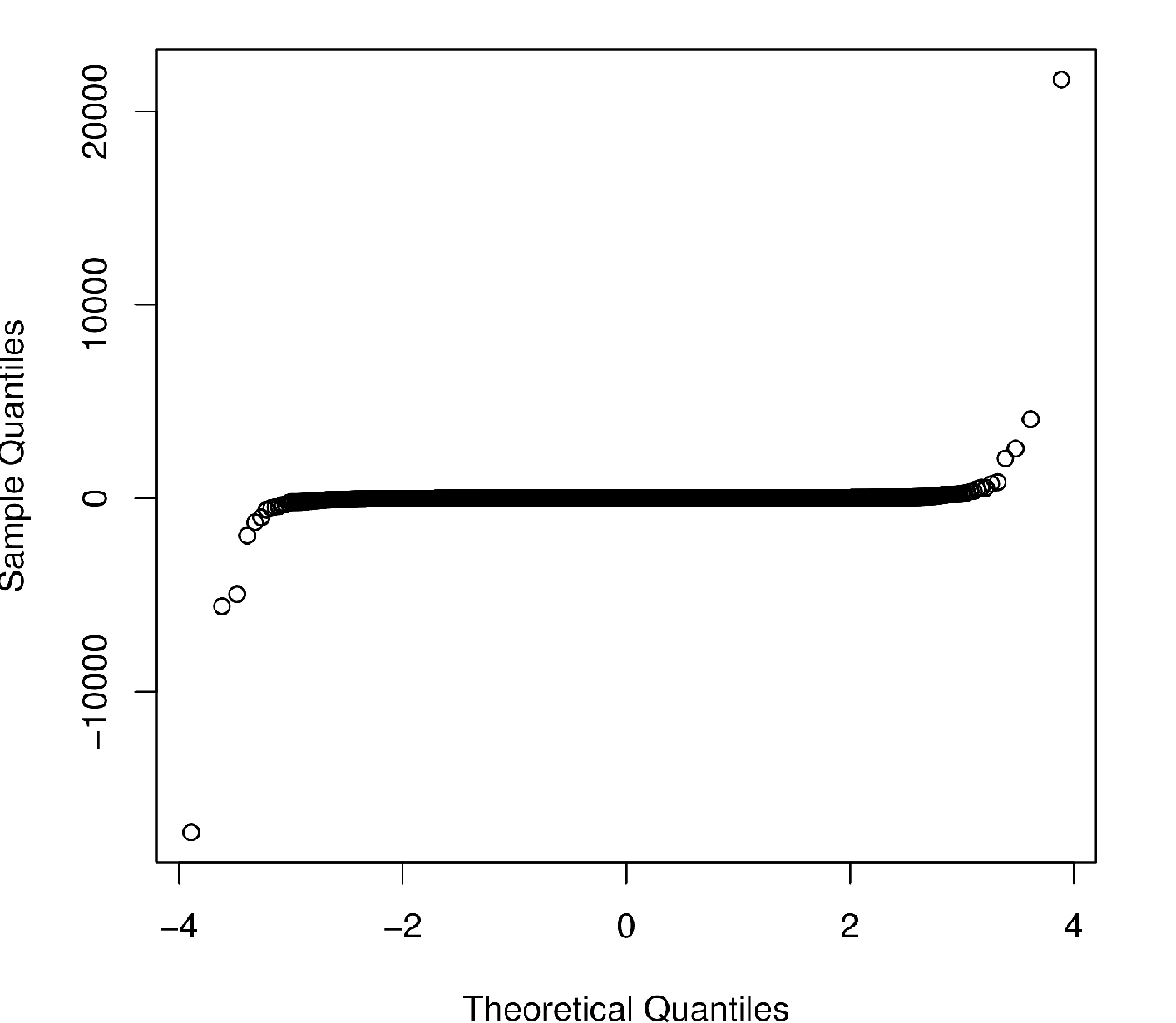}
\caption{The probability density function of the power law noise in the models that we use appear similar to Gaussian (normal) for values of α below 2 and more like a uniform probability density function for α above 2. This behavior can be displayed in the graph of the normal Quantile-Quantile (QQ) plot (upper panels), where a normal distribution would lie perfectly on the diagonal. This can be compared to the QQ plots of the uniform distribution (lower left) and Cauchy distribution (lower right). }
\label{qq}
\end{figure}

Gain fluctuations are an unfortunate but common feature of low noise amplifiers such as HEMTs \cite{Gallego2004,Wollack1995}.  Here we simulate the effect 
small gain fluctuations by imposing a small fraction of power law noise with $\alpha=1$ onto a white noise radiometer signal. The graph of frequency distributions of the power law random variable over time shows that estimates of the mean value have a greater dispersion than would be expected from the standard error of the mean.  The effect increases as a greater amount of gain fluctuation is inserted.  The S/N ratio no longer improves as fast as $1/\sqrt{\rm{time}}$.

\begin{figure}[ht!]
\centering
\includegraphics[width=1.15in,trim=225 204 215 200,clip]{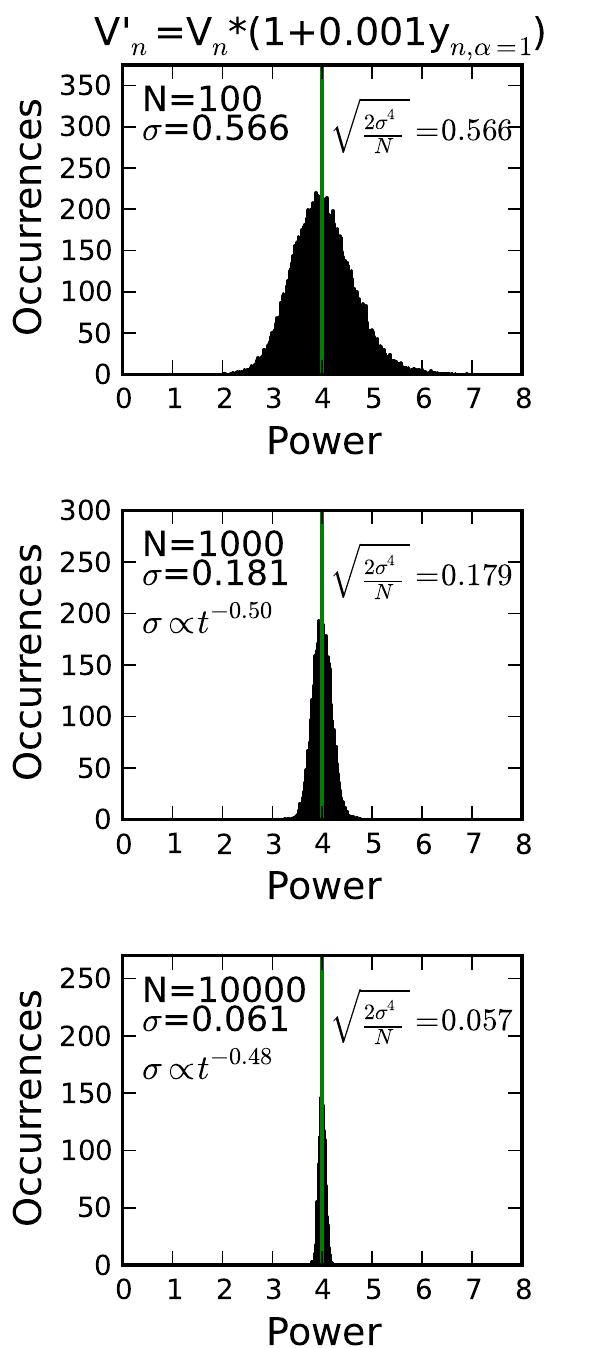}\hfill
\includegraphics[width=1.15in,trim=225 204 215 200,clip]{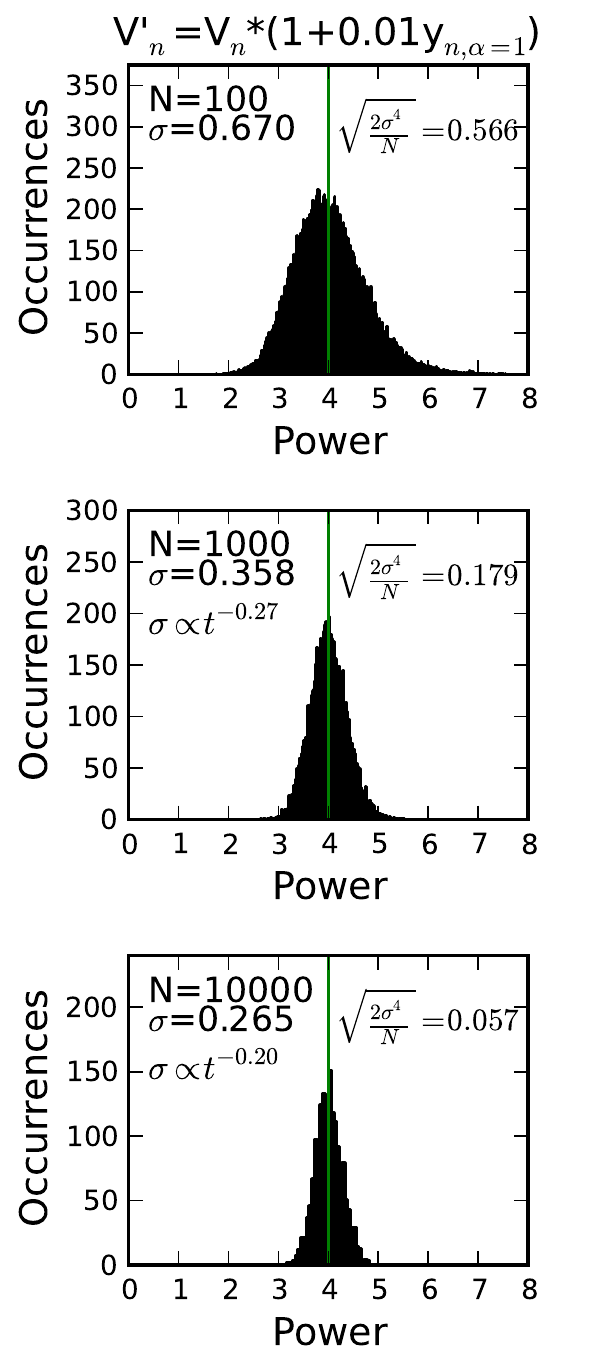}\hfill
\includegraphics[width=1.15in,trim=225 204 215 200,clip]{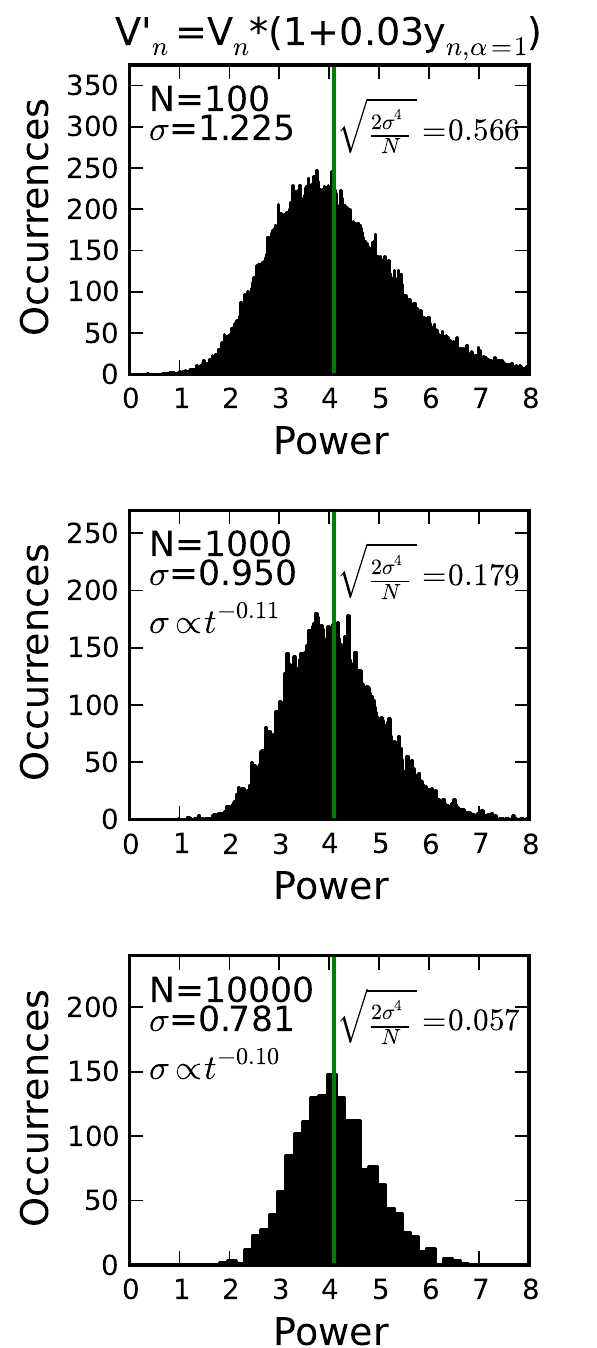}
\caption{Same as Fig.~1, except that the simulated signal contains an additional component of $\alpha=1$ noise imposed as gain fluctuations.  The amount of fluctuations increases from the left column to the right column.   The integration time of each sample increases from top to bottom.  As the gain fluctuations increase, the improvement in the S/N with longer integrations becomes less and the deviation of $\sigma$ from that predicted by the radiometer equation widens. }
\label{radiometeralpha}
\end{figure}

\section{Further reading}

A discussion of the radiometer equation in terms of single dish radio telescopes is given in the summer school lectures \cite{Campbell02}. Similarly, the application of the radiometer equation to cross-correlating interferometers is treated by several authors \cite{Wrobel99}, \cite{Crane89}, \cite{Thompson86}.

\section{Future work}
Our future work on this topic includes simulating the role of the radiometer equation in a two element interferometer, and examining the statistics of the digital data recorded by the Submillimeter Array (SMA) \cite{Blundell07,Hunter05} in Hawaii.





\bibliographystyle{ieeetr}
\bibliography{ms}

\begin{thebibliography}{10}

\bibitem{Johnson1928}
J.~B. {Johnson}, ``{Thermal Agitation of Electricity in Conductors},'' {\em
  Physical Review}, vol.~32, pp.~97--109, July 1928.

\bibitem{NIST2008}
J.~Randa, J.~Lahtinen, A.~Camps, M.~L. V. D.~M. Gasiewski, A.
  J.and~Hallikainen, M.~Martin-Neira, J.~Piepmeier, P.~W. Rosenkranz, C.~S.
  Ruf, J.~Shiue, and N.~Skou, ``Recommended terminology for microwave
  radiometry,'' {NIST Technical Note 1551}, National Institute of Standards and
  Technology, USA, August 2008.

\bibitem{Racette2005}
P.~{Racette} and R.~H. {Lang}, ``{Radiometer design analysis based upon
  measurement uncertainty},'' {\em Radio Science}, vol.~40, p.~5004, Oct. 2005.

\bibitem{Nyquist1928}
H.~{Nyquist}, ``{Certain Topics in Telegraph Transmission Theory},'' {\em
  Transactions of the A.I.E.E.}, pp.~617--644, Feb. 1928.

\bibitem{Fisher1930}
R.~A. {Fisher}, {\em {Statistical Methods for Research Workers}}.
\newblock {London}: {Oliver and Boyd}, 1930.

\bibitem{Oliver1965}
B.~M. {Oliver}, ``{Thermal and quantum noise},'' {\em IEEE Proceedings},
  vol.~53, pp.~436--454, May 1965.

\bibitem{Dicke}
R.~H. {Dicke}, ``{The Measurement of Thermal Radiation at Microwave
  Frequencies},'' {\em Review of Scientific Instruments}, vol.~17,
  pp.~268--275, July 1946.

\bibitem{Tsybulev2014}
P.~G. {Tsybulev}, M.~V. {Dugin}, A.~B. {Berlin}, N.~A. {Nizhelskij}, D.~V.
  {Kratov}, and R.~Y. {Udovitskiy}, ``{1/f-Type noise in a total power
  radiometer},'' {\em Astrophysical Bulletin}, vol.~69, pp.~240--246, Apr.
  2014.

\bibitem{Gallego2004}
J.~D. {Gallego}, I.~{L{\'o}pez-Fern{\'a}ndez}, C.~{Diez}, and A.~{Barcia},
  ``{Experimental results of gain fluctuations and noise in microwave low-noise
  cryogenic amplifiers},'' in {\em Noise in Devices and Circuits II}
  (F.~{Danneville}, F.~{Bonani}, M.~J. {Deen}, and M.~E. {Levinshtein}, eds.),
  vol.~5470 of {\em Society of Photo-Optical Instrumentation Engineers (SPIE)
  Conference Series}, pp.~402--413, May 2004.

\bibitem{Wollack1995}
E.~J. {Wollack}, ``{High-electron-mobility-transistor gain stability and its
  design implications for wide band millimeter wave receivers},'' {\em Review
  of Scientific Instruments}, vol.~66, pp.~4305--4312, Aug. 1995.

\bibitem{Campbell02}
D.~B. {Campbell}, ``{Measurement in Radio Astronomy},'' in {\em Single-Dish
  Radio Astronomy: Techniques and Applications} (S.~{Stanimirovic},
  D.~{Altschuler}, P.~{Goldsmith}, and C.~{Salter}, eds.), vol.~278 of {\em
  Astronomical Society of the Pacific Conference Series}, pp.~81--90, Dec.
  2002.

\bibitem{Wrobel99}
J.~M. {Wrobel} and R.~C. {Walker}, ``{Sensitivity},'' in {\em Synthesis Imaging
  in Radio Astronomy II} (G.~B. {Taylor}, C.~L. {Carilli}, and R.~A. {Perley},
  eds.), vol.~180 of {\em ASP Conference Series}, p.~171, 1999.

\bibitem{Crane89}
P.~C. {Crane} and P.~J. {Napier}, ``{Sensitivity},'' in {\em Synthesis Imaging
  in Radio Astronomy} (R.~A. {Perley}, F.~R. {Schwab}, and A.~H. {Bridle},
  eds.), vol.~6 of {\em Astronomical Society of the Pacific Conference Series},
  p.~139, 1989.

\bibitem{Thompson86}
A.~R. {Thompson}, J.~M. {Moran}, and G.~W. {Swenson}, {\em {Interferometry and
  synthesis in radio astronomy}}.
\newblock 1986.

\bibitem{Blundell07}
R.~{Blundell}, ``{The Submillimeter Array},'' {\em IEEE MTT-S International
  Microwave Symposium Digest}, pp.~1857--1860, July 2007.

\bibitem{Hunter05}
T.~R. {Hunter}, J.~W. {Barrett}, R.~{Blundell}, R.~D. {Christensen}, R.~S.
  {Kimberk}, S.~{Leiker}, D.~P. {Marrone}, S.~N. {Paine}, D.~{Cosmo Papa},
  N.~{Patel}, P.~{Riddle}, M.~J. {Smith}, T.~K. {Sridharan}, C.-Y.~E. {Tong},
  K.~H. {Young}, and J.-H. {Zhao}, ``{Dual frequency 230/690 GHz interferometry
  at the Submillimeter Array},'' in {\em Sixteenth International Symposium on
  Space Terahertz Technology}, pp.~58--63, May 2005.

\end{thebibliography}
%

%


\end{document}